\documentclass[12pt,preprint]{emulateapj}

\newcommand{\mymail}{oysteol@astro.uio.no}
 
\begin{document}

\altaffiltext{1}{Institute of Theoretical Astrophysics, University of Oslo, 
P.O. Box 1029 Blindern, N-0315 Oslo, Norway}
\altaffiltext{2}{Lockheed Martin Solar and Astrophysics Lab, 3251 Hanover St., 
Org. ADBS, Bldg. 252, Palo Alto, CA 94304, USA}
\altaffiltext{3}{Center of Mathematics for Applications, University of 
Oslo, P.O. Box 1053 Blindern, N-0316 Oslo, Norway}
\altaffiltext{4}{INAF-Osservatiorio Astrofisico di Arcetri, 
I-50125 Firenze, Italy}

\title{Search for high velocities in the disk counterpart of type II spicules.}

\author{{\O}. Langangen\altaffilmark{1}}
\email{\mymail}
\author{B. De Pontieu \altaffilmark{2}, M. Carlsson \altaffilmark{1,3}, V. H. Hansteen \altaffilmark{1,2,3}, G. Cauzzi \altaffilmark{4}, K. Reardon \altaffilmark{4}}

\begin{abstract}
  Recently, \citet{spic2} discovered a class of spicules that evolves
  more rapidly than previously known spicules, with rapid apparent
  motions of 50--150 km\,s${}^{-1}$, thickness of a few 100 km, and
  lifetimes of order 10--60 seconds. These so-called type II spicules
  have been difficult to study because of limited spatio-temporal and
  thermal resolution. 
  Here we use the IBIS
  instrument to search for the high velocities in the disk counterpart 
  of type II spicules.
  We have detected rapidly evolving events, with lifetimes that are
  less than a minute and often equal to the cadence of the instrument
  (19 secs).  These events are characterized by a Doppler shift that
  only appears in the blue wing of the Ca II IR line.  Furthermore the
  spatial extent, lifetime, and location near network, all
  suggest a link to type II spicules.  However, the magnitude of the
  measured Doppler velocity is significantly lower than the apparent
  motions seen at the limb. We use Monte Carlo simulations to show
  that this discrepancy can be explained by a forward model in which
  the visibility on the disk of the high-velocity flows in these
  events is limited by a combination of line-of-sight projection and
  reduced opacity in upward propelled plasma, especially in
  reconnection driven jets that are powered by a roughly constant energy
  supply.
\end{abstract}

\section{Introduction}
\label{sec:intro}

The solar chromosphere, when viewed at the limb, is dominated by a
variety of jet-like features that have been difficult to understand,
mostly because the spatio-temporal resolution of previous observations
has been insufficient to resolve much of the dynamics, which has led
to a multitude of different interpretations \citep{2000Sterling}.
Recently observations at high resolution (100--200 km) and cadence (a
few seconds) at the Swedish 1-m Solar Telescope \citep{2006fast} and with
the Solar Optical Telescope (SOT) onboard Hinode \citep{spic2} have
provided unprecedented views of the complex mix of highly dynamic and
finely structured features that dominate the chromosphere. A full
understanding of chromospheric energetics and coupling to the
transition region and corona requires a detailed study of what drives
this plethora of features and how they are connected to one another.

Some progress in this direction has been made by \citet{spic2} who
found that there are two fundamentally different types of spicules at
the limb, with very different dynamic properties. The type I spicules
evolves on timescales of order several minutes and seems to be driven
by shock waves that form when oscillations and flows leak into the
chromosphere along magnetic flux concentrations. These spicules seem
to correspond to dynamic fibrils (active regions) and a subset of
mottles (quiet Sun), when seen on the solar disk
\citep{2006Hansteen,2007DePontieu,2007Rouppe}.  Hinode/SOT
observations revealed a second class of spicules (``type II'') that
occur on timescales of just a few tens of seconds during which they
are seen to rise and then rapidly disappear, possibly because they are
heated out of the bandpass. These newly discovered
features, which dominate the limb in coronal holes and (to a lesser
extent) the quiet Sun \citep{spic2}, remain mysterious since the
line-of-sight projection makes it difficult to determine whether the
high apparent velocities (50--150 km\,s$^{-1}$) are associated with
real plasma flows of the same magnitude. There is a significant literature on 
the properties and relationship of mottles and spicules in various wavelengths. 
However, the enormous advances in spatial and temporal resolution of the Hinode 
data, and the lack of high velocity observations in the classical observations 
limit the usefulness of detailed comparisons with previous observational 
reports.

In this letter we present a candidate for the disk counterpart of type II
spicules. We observe short lived blue-shifted excursions in the 
line profile of the Ca II 854.2~nm line that are not succeeded by an
obvious redshift. 
Similar events have been reported
earlier \citep[e.g][]{1998Wang,1998Chae}, but on temporally and
spatially more extended features.  To explain the differences observed
between the limb and the disk we present numerical experiments using
radiative transfer calculations and Monte Carlo simulations.

\section{Observations}
\label{sec:obs}

\begin{figure}[!ht]
\includegraphics[width=0.5\textwidth]{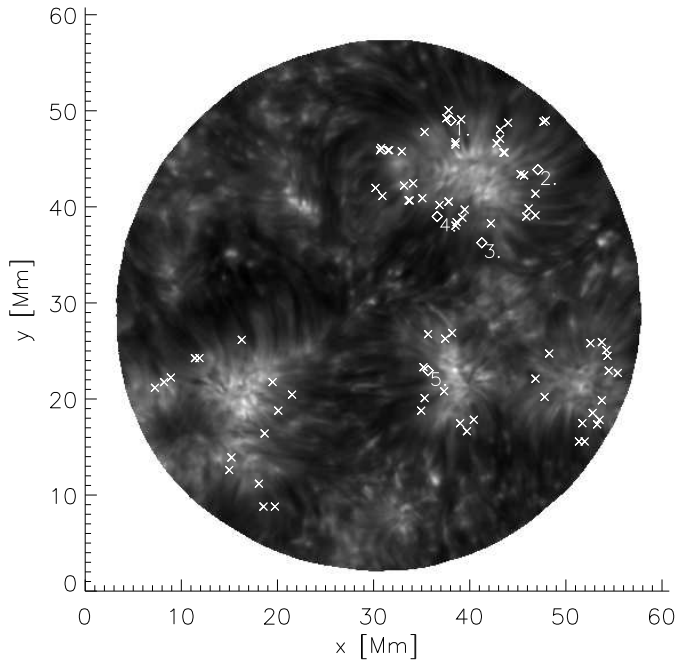}
\caption{
  A snapshot of a line center image showing an overview of the field of view.
  Crosses indicate the positions of the observed rapid blueshifted
  excursions, with the five examples seen in Fig.\ref{spectra} marked
  as diamonds with the corresponding number.  }
\label{overview}
\end{figure}

\begin{figure*}[!ht]
\includegraphics[width=\textwidth]{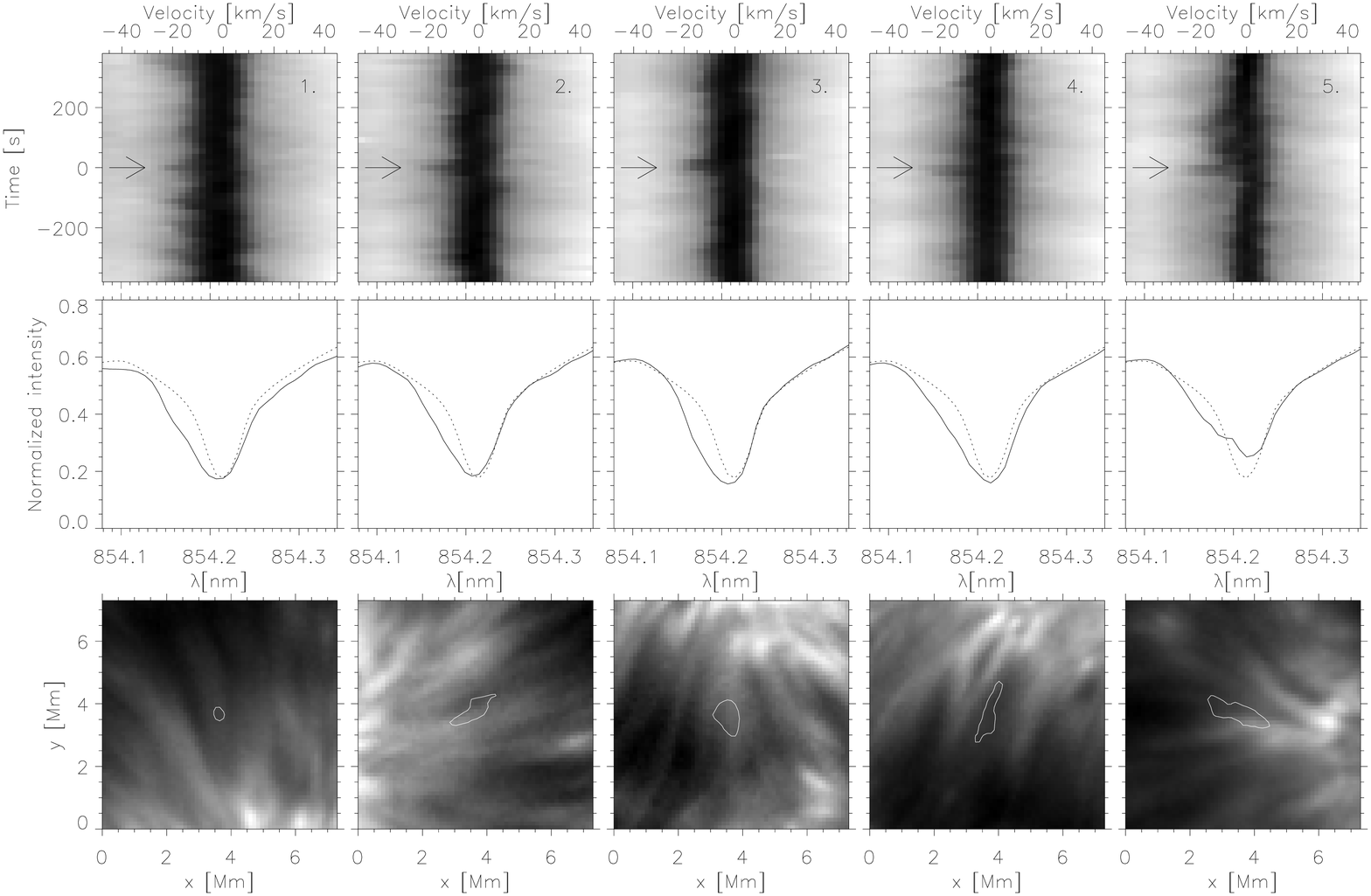}
\caption{Five examples of RBEs are shown.
  The top row shows the temporal evolution of the RBE (arrow). 
  The detailed spectra (central row) of the RBEs (solid line) 
  is shown together with the mean profile (dotted line). 
  The line center image is shown in the bottom
  row.  The contour (white line) shows the spatial extent of the RBEs. 
  Note the alignment of the RBEs with the field lines, see also
  Fig.\ref{overview}. }
\label{spectra}
\end{figure*}
The observations were obtained using the Interferometric BIdimensional
Spectrometer (IBIS) mounted at the Dunn Solar Telescope (DST) of the
National Solar Observatory. For a more thorough description of the
instrument see \citet{2006Cavallini}. The data used in this letter
was obtained on 2 June 2004, during a period of good to excellent 
seeing conditions lasting for about 55 minutes (UT 14:59--15:54). 
For a detailed description of the data set and the reduction routines
we refer the reader to other work using the same data set
\citep[e.g.][]{2008Cauzzi}.  The data was obtained with a scanning
sequence of 27 steps in the Ca II 854.2~nm line, with a step size
of 8--16 pm and a spectral transmission
FWHM of 4.4 pm. This scan covers a spectral region of -0.12~nm to
about 0.16~nm from line center. The signal-to-noise ratio is about 130 in 
the linewing. The diffraction limit of the
telescope at 854.2~nm is $\frac{\lambda}{D}=0.23$\arcsec{} and the
pixel size is 0.082\arcsec{} but with $2\times2$ binning the effective 
size is 0.165\arcsec{}. The scanning of the Ca line takes
7 seconds and was repeated every 19 seconds pointing at an enhanced 
network very close to the disk center. The field of view (FOV) covers four
rosette structures (see Fig.\ref{overview}).

\section{Data Analysis}\label{sec:res}

Visual inspection of the data set shows that the general dynamic
behavior of the spectral line varies dramatically across the different
regions in the FOV \citep[e.g. network, internetwork (IN), and canopy, see]
[]{2007Vecchio}.  
The IN is cluttered with brightenings
caused by the 3 min shocks as explained by \citet{1997Carlsson}, while
the center of the rosettes are covered with mottles \citep[with a
spectral signature very similar to that of dynamic
fibrils,][]{2008Langangen}.  At the edges of the rosettes
we observe a marked difference in the dynamic behavior of the line
profile.
The line is wider and there are no dominating shocks, as compared to
the network and the IN. It is in this region, at the edge of
the rosettes that we find the blue shifted excursions (see Fig.\ref{overview}
and Fig.\ref{spectra}).

The events have been identified in a semi-automatic fashion.  First we
perform a search for changes in the intensity on the blue side of the
line at around 20 km\,s${}^{-1}$.  We then exclude all events that
show a succeeding redshift, since these kinds of events are identical
to the temporal evolution of the velocity in a dynamic fibril or an IN
shock.  Finally we exclude all events that occur during times of worse
seeing quality.  Such an automatic search results in a manageable
number of events. The events found by the search routine were manually
inspected and only clearly blueshifted and short lived events were
chosen for further analysis.  In this fashion we identified 87 blue
shifted events distributed over the whole time series, 
from now on denoted as rapid blueshifted excursions (RBE).
We have done a visual search for separate downflows associated with the RBEs 
and have found no clear candidates.

To measure the spatial extent of the RBEs we calculate the spatial
correlation of the dynamic behavior of the line profile at the
semi-automatically flagged location with that of surrounding pixels.
We choose the most pronounced timestep in the RBE as anchor, subtract
the mean line profile from the spectra, and calculate 
the correlation coefficient of the temporal evolution of the blue side of
the RBE between neighboring pixels. 
The area covered by pixels with higher than 80
$\%$ correlation to the anchor pixel is used as a measure of the
spatial extent of the structure. 
The RBEs can be elongated ($78\%$), round ($8\%$), or 
have an irregular shape ($14\%$), see Fig.\ref{spectra}.
The length of the elongated RBEs is usually 0.5--1.5~Mm with an average
of 1.2~Mm and the width is usually 0.3--0.6~Mm with an average of 0.5 Mm. 
The small round RBEs are often not more than 0.5 Mm in diameter. 
It is likely that some of the elongated shape of the RBEs is caused by the 
smearing due to the long scanning times (7 secs) of the data. 

From the measured RBEs we derive a mean lifetime of 45~$\pm$~13~seconds.  
It is difficult to accurately measure plasma velocities in
the RBEs, since many of the spectral profiles seem to consist of more
than one component (but without any separation between components).
However, a rough estimate of the velocity based on visual inspection
and comparison to radiative transfer calculations suggests velocities
of order 15--20 km\,s${}^{-1}$ (see \S \ref{sec:num} for more
discussion of this topic).

\section{Numerical experiments}\label{sec:num}

The similarities in properties such as uniquely upward motion, rapid
disappearance, lifetime, and location suggest a link between type II
spicules and RBEs. However, there is an apparent lack of very high
velocity RBEs, especially compared to the 50--150 km s$^{-1}$
velocities seen in type II spicules. To investigate these differences
in the observed properties of type II spicules and RBEs we perform
several numerical experiments.  First we construct a toy model for the
RBEs: a 100 km wide component with a LOS velocity of 15 and 20 km/s is
introduced in a FAL~C model \citep{1993Fontenla}, 
at a variety of different heights. We then
use MULTI \citep{multi} to calculate the impact of these higher
velocity components on the Ca II 854.2~nm line profile (as seen on the disk).  
While this may be unrealistic, it can give some insight into the effect of 
atmospheric motions on the line profile. A full understanding of 
the radiative transfer in these events will require a more sophisticated 
approach e.g. a cloud model \citep{1994Heinzel}, or more likely 
non-LTE radiative transfer from a 3D MHD model. 
This numerical experiment indicates that events that occur at the top of the 
chromosphere (where lower densities occur) are increasingly more difficult 
to observe. 

\begin{figure*}[!ht]
\includegraphics[width=\textwidth]{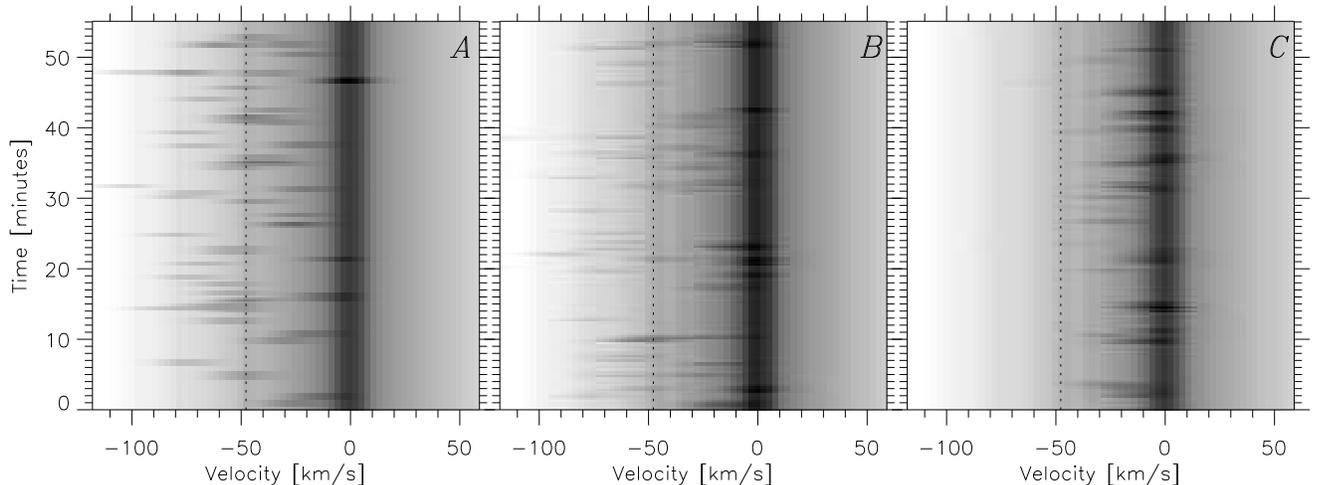}
\caption{
Results from the Monte Carlo simulations;
a simulation with infinite scale height  (panel A), 
a simulation with a scale height of 1500 km (panel B), 
and a simulation with a scale height of 1500 km and
random height of energy input panel C) are shown. 
The vertical dashed line is the limit of the IBIS observations. 
}
\label{mc}
\end{figure*}

As a next step we investigate whether line-of-sight projection is
behind the lower velocities (15--20 km s$^{-1}$) observed in RBEs on
the disk compared to the high velocities of type II spicules at the limb. 
Perhaps the strong inclination from the vertical of the magnetic field lines
at the edge of rosettes causes a reduction in observed velocities
along the line-of-sight? For this purpose we perform a Monte Carlo
simulation in which 50 type II spicules are allowed to occur
(randomly) during a time period of 55 min (at 19 s cadence).  Each
spicule has a lifetime randomly chosen from a Gaussian distribution
around 40 s with a standard deviation of 20 s. During their lifetime,
the spicules are assigned an upward velocity chosen from a Gaussian
distribution around 70 $\pm$ 15 km\,s${}^{-1}$. The spicules have
random orientation with a uniform
distribution of angles between the spicule axis and local vertical
between 0${}^{\circ}$ and 90${}^{\circ}$. We then synthesize the
impact of these spicules on the Ca II 854.2~nm line by assuming each
type II spicule is associated with a Gaussian absorption component
with a depth of 10$\%$ of the continuum value, with a half-width of 
10~km\,s${}^{-1}$. The result of such a simulation (Fig.\ref{mc} A) shows a 
large number of high velocity events with Doppler shifts larger than 30
km/s, which we do not see in the IBIS data. We should note that
assigning spicules orientations that are more inclined from the
vertical (e.g., as in the inclined regions at the edge of rosettes)
does lead to a reduction of high-velocity events in the simulated IBIS
Dopplergrams. However, Hinode/SOT data shows that many type II
spicules are more vertical, so 
a skewed angle distribution cannot be the reason for the 
mismatch in velocities between type II spicules and RBEs.

Next we investigate whether the decreasing density with height in
spicules could lead to a reduced visibility of nearly vertical
spicules in which the density rapidly decreases with height. A simple
way of incorporating this effect is to decrease the absorption caused
by each spicule as a function of height, using the observed intensity scale
height \citep{spic2}. To test the maximal effect we use the lower observed 
scale height of 1500 km. The absorption component is
then allowed to decrease in strength according to the fraction of the scanning 
time (7 secs) and the cadence (19 secs) the spicule occupies at the different
heights. Highly inclined events (e.g., at the edge of the rosettes) do
not reach high up, so do not suffer much attenuation of the default
absorption. An example of such a simulation (Fig.\ref{mc} B) shows that the
visibility of events with high Doppler shifts
is reduced significantly. However, it appears there are still a larger
number of such events than we observe in the IBIS data.

A resolution to this problem comes when we assume that the spicules
are caused by a local release of energy, such as in a reconnection
event. If the
reconnection events occur over a range of heights and each event is
assumed to release a roughly constant amount of energy, then the
dropoff in density with height will cause events that
occur in the upper chromosphere to naturally lead to spicules with
lower density and higher velocity than events driven by reconnection
lower down. We incorporate this idea into the Monte Carlo simulations
by allowing each spicule to be driven by reconnection at a random
height chosen from a uniform distribution between 1000--2000 km (in
the FAL~C model). The velocity of the spicule caused by the
reconnection is then assumed to be given by
$v=\sqrt{2U_k/\rho}$, where $U_k$ is the kinetic energy density in the
event, and $\rho$ is the density in the FAL~C model at the
height of the event. The energy density of the event is now a free
parameter, but there are a whole range of values for this parameter
(10--70 ergs\,cm${}^{-3}$) that produce simulated Dopplergrams
that are very similar to the RBEs in the IBIS data. One example of
such a simulation with $U_k=25$ ergs\,cm${}^{-3}$ is shown Fig.\ref{mc} C.  
While this simulation produces a whole range of high velocity
events, these are not visible against a bright background, and only
visible at the limb (as type II spicules) where there is no background
radiation. This can explain the lack of high velocity RBE's on the
disk, and the predominance of high velocity events at the limb (where
there is a bias towards mostly vertical high velocity events that
reach greater heights). 

\section{Conclusions}\label{sec:con}

We have presented a candidate for the disk counterpart of type II
spicules: rapid blue-shifted events (RBEs). The 
RBEs are found 
near the edges of the network, where movies show that straw-like features 
are present as well. The lifetimes of the RBEs are about 45 seconds,
which is very close to those of type II spicules (40 s). The length of
the elongated RBEs is about 1.2~Mm and the width is 0.5~Mm.
The length of the RBEs is shorter than the type II spicules, which
might be caused by the intrinsic differences in visibility between the
two kind of observations.  Since RBEs are identified by a Dopplershift
they are associated with mass motion. The magnitude of the mass motion
in RBE's 
is lower than the apparent motion observed in the
Type II spicules.  Monte Carlo simulations suggest that this
discrepancy is to be expected if there is an inverse relationship
between density and velocity of the events: high density, low velocity
events show enough absorption to be visible in IBIS data taken on the
disk, whereas low density, high velocity events are more clearly
visible at the limb (since they reach greater heights), but have
little visibility on the disk because they have little opacity. Such
an inverse relationship would be expected if type II spicules are
caused by reconnection events that occur over a range of heights in
the chromosphere, and in which the amount of energy is roughly
constant from event to event. Such a scenario is not unrealistic, and
is supported by the presence of RBEs at the edge of rosettes where the
relentless magnetoconvection pushes the ubiquitous IN flux
into the strong network flux concentrations. Compounding factors that
help explain the mismatch in observed velocities include the effects
of variable LOS velocity due to the magnetic field topology and the
fall in intensity due to the decrease in density as the spicules
reaches greater heights. We show that a combination of these factors
can explain the visibility of RBEs on the disk and type II spicules at
the limb.
  

\bibliographystyle{apj}
\bibliography{ms}

\end{document}